\documentclass[prl,twocolumn,showpacs, amsmath,amssymb]{revtex4} 
\begin{document}
\def\be{\begin{equation}}
\def\ee{\end{equation}}
\def\bc{\begin{center}} 
\def\ec{\end{center}}
\def\bea{\begin{eqnarray}}
\def\eea{\end{eqnarray}}
\newcommand{\avg}[1]{\langle{#1}\rangle}
\newcommand{\Avg}[1]{\left\langle{#1}\right\rangle}

\title{The entropy  of network ensembles}

\author{Ginestra Bianconi}

\affiliation{The Abdus Salam International Center for Theoretical 
Physics, Strada Costiera 11, 34014 Trieste, Italy }
\abstract{
In this paper we generalize the concept of random networks to describe
networks with non trivial features by a  statistical mechanics approach.
 This framework is able to describe ensembles of undirected, directed
 as well as  weighted networks. These networks might have not trivial community
 structure or, in the case of networks embedded in a given space, non trivial
 distance dependence  of the link probability.  These ensembles are
 characterized by their entropy which evaluate the cardinality of
 networks in the ensemble.
The general framework  we present  in this paper is able to describe microcanonical ensemble of networks
as well as canonical or hidden variables network ensemble with
significant implication for the formulation of network constructing
algorithms.
 Moreover in the paper  we define and 
 and characterize in particular    the {\em structural entropy},
 i.e. the entropy of the ensembles of undirected uncorrelated simple
 networks with given degree sequence. We  discuss   the
 apparent paradox that scale-free degree distribution are
 characterized by having small structural entropy but are so widely
 encountered in natural, social and technological complex systems.We
 give the proof that while scale-free networks ensembles have small structural
 entropy, 
they also correspond to the most likely degree distribution with the
corresponding value of the structural entropy.
} 
}

\maketitle
\section{Introduction}

The quantitative measure of the order present in complex systems and the possibility to extract information from the complex of interactions in cellular, technological and social networks is a topic of key interest in modern statistical
mechanics.
The field of complex networks \cite{Dorogovtsev,Latora} has having a
rapid development and a large success in this respect due to   the
wide applicability of simple concepts coming from graph theory. The
characterization of the structure of different networks has allowed
the scientific community to compare systems of very different nature. 
Different statistical mechanics tools have been devised to describe  the different level of organization of  real
networks. A description of the structure of a complex network  is presently performed by measuring different quantities as
 (i) the density of the links, (ii) the degree sequence \cite{BA}, (iii)
the degree-degree correlations \cite{Vespignani_corr,Sneppen,Berg}, (iv) the
clustering coefficient \cite{SW,Modular}, (v) the k-core structure \cite{k-core_kirk,k-core_doro,k-core_vesp}, (vi) the community structure \cite{Newman1,Danon,Newman2,Latora}
and finally the nature of the embedding space
 \cite{Kleinberg,Dorogovtsev_new, Boguna2}.
Moreover, is the network is weighted,  strength/degree correlations
\cite{Vespignani_wei} 
and if the network is directed,   in-degree/out-degree correlations
\cite{loop_adilson} are significant characteristics of the network
 These phenomenological quantities describe the local or non-local topology of the
 network and do affect dynamical models defined of them
 \cite{Dorogovtsev}.

While many different  statistical mechanics models have been proposed
\cite{Burda_stat,Doro_stat,hv1,hv2,hv3,hv4,Boguna_hv,Garlaschelli} to
describe how the power-law degree distribution can arise in complex
networks, little work has been done on the problem of measuring the
level of organization and "order" in the frame of theoretical statistical mechanics. 
Only recently, in the field of complex networks attention has been
addressed to the study of entropy measures 
\cite{Burda_en,Opt,bauer,entropy,Ldev,Latora2} able to approach this problem.
In \cite{entropy} the {\it entropy of a given ensemble} as
the normalized logarithm of the number of networks in the
ensemble has been introduced. This quantity can be used to asses the
role that a given structural characteristics have in shaping the network.
 In fact, given a real network, a subsequent series of randomized
 networks ensembles can be build each subsequent ensemble sharing one
 additional structural characteristic with the given network.
The entropy of these subsequent networks ensembles would  decreases
as we proceed adding constraints and
 the difference between the entropies in two subsequent ensembles quantifies how restrictive is the introduced additional
constraint.
In the first part of this  paper construct a general statistical
mechanics framework for the construction of generalized random network
ensembles which satisfy given structural constraints. We call these
ensembles "microcanonical". We also
describe how to construct "canonical" network ensembles or  generalized  hidden variable
\cite{hv1,hv2,hv3,hv4,Boguna_hv,Garlaschelli} models.
Subsequently we make an account of most of the network ensembles that can be formulated:  the ensemble
of undirected networks with given number of links and nodes, the ensemble of
undirected networks with given degree sequence, with given  spatial
embedding and community structure. Some of these network ensembles
where already presented in \cite{entropy} and we report their
derivation here for completeness. This approach is further extended
to weighted networks and directed networks.
Finally  we  focus our attention on the structural entropy, i.e. the
entropy of an ensemble of uncorrelated undirected simple networks of given degree
sequence. The structural entropy of a power-law network with constant average
degree is monotonically decreasing as the power-law
exponent $\gamma\rightarrow 2$.
This result could appear in contradiction with the wide occurrence of
power-law degree distribution in complex networks.
Here we show by a statistical mechanics model that scale-free degree
distribution are  the most likely degree distribution at given
small value of structural entropy while  Poisson degree
distributions are the most likely degree distribution of networks with
maximal structural entropy.

This result indicates that the scale-free degree distributions emerges
naturally when considering networks ensembles with small structural
entropy and therefore larger amount of order.

The appearance of the power-law degree distribution reflects
 the tendency of social, technological and especially biological networks toward ``ordering''.
This tendency is at work regardless of the mechanism which is
driving their evolution that can be either a
preferential attachment mechanism \cite{BA}, or a ``hidden variables''
mechanism \cite{hv1,hv2,hv3,hv4,Boguna_hv,Garlaschelli} or  some other statistical
mechanics mechanism \cite{Burda_stat,Doro_stat}.

\section{Statistical mechanics of network ensembles}

A network of $N$ labeled nodes $i=1,2,\ldots, N$ is uniquely defined
by its adjacency matrix ${\bf a}$ of matrix elements $a_{ij}\ge 0$ with
$a_{ij}>0$ if and only if 
there is a link between node $i$ and node $j$.
Simple networks are networks without tadpoles or double links, i.e.  $a_{ii}=0$ 
and $a_{ij}=0,1$. Weighted networks describe heterogeneous
interactions between the nodes and the matrix elements $a_{ij}$ can
take different null or positive values, while directed networks are
described by non-symmetric adjacency matrices ${\bf a}\neq {\bf a^T}$
where we have indicated by ${\bf a^T}$ the transpose of the matrix ${\bf
a}$.

A structural constraint on a network can always be formulated as a
constraint on the adjacency matrix of the graph, i.e.
\be
{\vec F}({\bf a})=\vec{C}.
\ee

In order to describe "microcanonical" network ensembles  with given structural
constraints in \cite{entropy} and in the following we will use a
statistical mechanics perspective. Therefore we  define a partition
function $Z$ of the ensemble in the following way
\be
Z=\sum_{{\bf a}} \delta\left[\vec{F}({\bf
    a})-\vec{C}\right]e^{\sum_{ij} h_{ij}\Theta(a_{ij})+r_{ij}a_{ij}}
\label{Z0.eq}
\ee
where, for simplifying the problem ,$\vec{F}({\bf a})$ and ${a_{ij}}$
take only integer values, and   $\delta[\cdot]$ indicate the Kronecker
delta and $\Theta(x)=1 $ if $x>0$ and $\Theta(x)=0$ if $x=0$. Moreover, in $(\ref{Z.eq})$, the auxiliary fields $h_{ij}$ have been introduced as in classical statistical
mechanics.
The entropy  per node  $\Sigma$ of the network ensemble is defined as 
\be
\Sigma=\frac{1}{N}\left.\ln(Z)\right|_{h_{ij}=r_{ij}=0\, \forall \, (i,j)}.
\ee
The marginal probability for a certain value of the element $a_{ij}$
of the adjacency matrix is given by
\be
\pi_{ij}(A)=\frac{1}{Z}\sum_{{\bf a}}\delta(a_{ij}-A)\delta\left(\vec{F}({\bf    a})-\vec{C}\right).
\label{piij}
\ee
The probability of a link $p_{ij}$   is given by 
\be
p_{ij}=\left. \frac{\partial \ln Z}{\partial
    h_{ij}}\right|_{h_{ij}=r_{ij}=0 \,\forall \,(i,j}.
\label{pij}
\ee
In and ensemble of  weighted network we can define also the average weight $w_{ij}$ of a link
between node $i$ and node $j$ as equal to
\be
w_{ij}=\left.\frac{\partial \ln Z}{\partial
    r_{ij}}\right|_{h_{ij}=r_{ij}=0\, \forall\, (i,j)}
\ee 
In a ``microcanonical'' network ensemble all the networks that satisfy
a given structural constraint have equal probability.
Therefore  the probability of a network $G$, described by the adjacency
matrix ${\bf a}$, is given in  the ``microcanonical'' ensemble  by
\be
P_M({\bf a})=e^{-N\Sigma} \delta\left[\vec{F}({\bf
    a})-\vec{C}\right]
\ee
If we allow for "soft" structural constraints in network ensemble we
can describe "canonical" network ensemble. 
In a "canonical'' network ensemble each network ${\bf a}$ has a different
probability given by
\be
P_C({\bf a})=\prod_{ij} \pi_{ij}(a_{ij})
\label{CW}
\ee
expression that for ensemble of simple networks take the form
\be
P_C({\bf a})=\prod_{ij} p_{ij}^{a_{ij}}(1-p_{ij})^{1-a_{ij}}.
\ee
If the link probabilities $\pi_{ij}(a_{ij})$ are chosen equal to  $(\ref{piij})$
and $(\ref{pij})$, then we have that the structural constraints $\vec{F}({\bf a})=\vec{C}$
are satisfied in average, i.e.
\be
\avg{\vec{F}({\bf a})}_{P_C({\bf a})}=\vec{C}
\ee
where the average $\avg{\cdot}_{P_C({\bf a})}$ indicates the average
over the canonical ensembles $(\ref{CW})$.
The statistical mechanics formulation of network ensemble is always
well defined. For  network structural constraints that do not  correspond to
feasible networks \cite{Toroczkai}  the entropy of the network
ensemble  is  nevertheless $\Sigma=-\infty$.
Although the definition of the statistical mechanics problem is
always well
defined, the calculation of the partition function by saddle point
approximation can only be performed if the number of constraints
${F}_{\alpha}$ with  $\alpha=1,\ldots ,M$ is at most extensive, i.e
$M={\cal O}(N)$. In addition to that, in the paper we are going to consider only
linear constraints on the adjacency matrix. Further developments on
this statistical mechanics framework  will
involve pertubative approach to solve non linear structural
constraints.

\section{ Undirected simple networks}
 
In an undirected simple network the adjacency matrix elements
are zero/ one ($a_{ij}=0,1$) and the tadpoles are forbidden( $a_{ii}=0
\, \forall i$).
We can consider for these networks different types of structural constraints. In
the following we list few of them of particular interest.
\begin{itemize}
\item
{\it i)}
The  ensemble $G(N,L)$ of random networks 
with given number of nodes
$N$ and links $L=\sum_{i<j}a_{ij}$ 
(providing  in this way a statistical
mechanics formulation of the   $G(N,L)$ random ensemble).
In this case we have the structural constraint
\be
\vec{F}({\bf a})-\vec{C} = \sum_{i<j} a_{ij}-L=0
\ee
\item
{\it ii)} 
The configuration model, i.e. the ensemble of networks with given degree sequence
 $\{k_1,\dots,k_N\}$ with $k_i=\sum_j a_{ij}$.
In this case the structural constraints are given by
\bea
{F}_{\alpha}({\bf a})-{C}_{\alpha}=\sum_{j}
a_{{\alpha}j}-k_{\alpha}=0
\label{conf.eq}
\eea 
for $\alpha=1,\ldots, N$.
\item
{\it iii)} 
The network with given degree sequence
$\{k_1,\dots,k_N\}$ and given
average nearest neighbor connectivity
$k_{nn}(k)=[\sum_{i,j}\delta(k_i-k) a_{ij} k_j]/(kN_k)$ of nodes of degree
$k$ (with $N_k$ indicating the number of nodes of degree $k$ in the network).
In this case the structural constraints are given by
\bea
{F}_{\alpha}({\bf a})-{C}_{\alpha}=\sum_{j} a_{{\alpha}j}-k_{\alpha}=0 
\label{dc1.eq}\eea
for $\alpha=1,\ldots,N$ and
\be
{F}_{\alpha}({\bf a})-{C}_{\alpha}=\sum_{ij}\delta(k_i-k) a_{ij} k_j-kN_k
k_{nn}(k)
\label{dc2.eq}\ee
for $\alpha=N+1,\ldots N+K.$ 
with $K$ indicating the maximal connectivity of the network.
\item
{\it iv)}
The network ensemble with given degree sequence and given
community structure. For these network we assume  that 
 each node is assigned a feature 
$\{q_1,\ldots q_N\}$ and we fix the number of links between nodes of
different features $A(q,q^{\prime})=\sum_{i<j}\delta(\underline{q_{ij}}-q)
\delta(\overline{q_{ij}}-q^{\prime})a_{ij}$ with
$\underline{q_{ij}}=\min(q_i,q_j)$ and $\overline{q_{ij}}=\max(q_i,q_j)$.
In this case the structural constraints are given by
\be
{F}_{\alpha}({\bf a})-{C}_{\alpha}=\sum_{j} a_{\alpha j}-k_{\alpha}=0 
\label{c1.eq}\ee for $\alpha=1,\ldots,N$ and 
\bea
{F}_{\alpha}({\bf a})-{C}_{\alpha}&=&
\sum_{i<j}\delta(\underline{q_{ij}}-q)
\delta(\overline{q_{ij}}-q^{\prime})a_{ij}
\nonumber \\
&&-A(q,q')
\label{c2.eq}
\eea
for $\alpha=N+1,\ldots N+Q(Q+1)/2$
with $Q$  equal to the number of different features of the nodes.
Here an in the following in order to have an extensive number of
constraints we assume $Q={\cal O}( N^{1/2})$.
\item
{\it v)} 
The  ensemble of networks with given degree sequence and dependence of
the link probability on the distance of the nodes  in an embedding
 geometrical space. In this ensemble  we consider fixed
spatial distribution of nodes in space $\{\vec{r}_1,\ldots,\vec{r}_N\}$
and we consider all the networks compatible with the given degree
sequence and the number of links linking nodes in a given distance
interval. Therefore we take ${\Lambda}$ distance intervals
 $I_{\ell}=[d_{\ell},d_{\ell}+(\Delta d)_{\ell}]$ with $\ell=1,\dots, \Lambda$, and
we fix the number of links linking nodes in a given distance interval.
The structural constraint involved therefore the vector  $B(d_{\ell})=\sum_{i<j}\chi_{\ell}(d_{i,j}) a_{ij}$
where $d_{ij}=d(\vec{r}_i,\vec{r}_j)$ is the distance between node $i$
and $j$ in the embedding space and the characteristic function
$\chi_{\ell}(x)=1$ if $x\in[d_{\ell},d_{\ell}+(\Delta d)_{\ell}]$ and $\chi_d(x)=0$ otherwise.
In this case the structural constraints can be expressed as
\be
{F}_{\alpha}({\bf a})-{C}_{\alpha}=\sum_{j} a_{{\alpha}j}-k_{\alpha}=0
\label{d1.eq}
\ee for  $\alpha=1,\ldots,N$
and 
\be
{F}_{\alpha}({\bf a})-{C}_{\alpha}=\sum_{i<j}\chi_{\ell}(d_{i,j})
a_{ij}-B(d_{\ell})
\label{d2.eq}
\ee
for $\alpha=N+1,\ldots N+\Lambda$.

\end{itemize}

\subsection{The  G(N,L) and the G(N,p) ensembles}
The networks in the $G(N,L)$  ensemble have  given number of nodes $N$ and
links $L$.
The entropy of this ensemble  is given by the logarithm of the binomial
\bea
N\Sigma_0=\left(\begin{array}{c} \frac{N(N-1)}{2}\\ L\end{array}\right)
\eea
(we always assume  distinguishable nodes in the networks \cite{Burda_en}).
The probability $p_{ij} $ of a given link $(i,j)$ is given by
$p_{ij}^{(0)}={L}/({N(N-1)/2})$ for every couple of nodes $i,j$.
The ensemble $G(N,p)$ is the "canonical" ensemble corresponding to the
"microcanonical" $G(N,L)$ ensemble.

\subsection{The configuration ensemble}

In the configuration ensemble we consider all the networks with given
degree sequence.
Using $(\ref{Z0.eq})$ $(\ref{conf.eq})$ the partition function of the ensemble can
be explicitly written as 
\be
{Z}_1=\sum_{\{a_{ij}\}}\prod_i\delta(k_i-\sum_j a_{ij})e^{\sum_{i<j}h_{ij}a_{ij}}
\ee 
Expressing the delta's in the integral form with Lagrangian multipliers
$\omega_i$ for every $i=1,\dots N$ we get
\be
{Z}_1=\int {\cal D}\omega \ e^{-\sum_i\omega_i k_i}\prod_{i<j}\left(1+e^{\omega_i+\omega_j+h_{ij}}\right)
\ee
where ${\cal D}\omega=\prod_i d\omega_i/(2\pi)$.
We solve this integral by saddle point equations accounting also for
 second order terms of the expansion.
The entropy of this ensemble of networks {can be approximated in the
large network limit $N\gg 1$ with}
\bea
N\Sigma_1^{und}&\simeq&-\sum_i\omega_i^*k_i+\sum_{i<j}\ln(1+e^{\omega_i^{\star}
 +\omega_j^{\star}})\nonumber \\
&&-\frac{1}{2}\sum_i \ln(2\pi\alpha_i)
\label{Sigma_1}
\eea
with the Lagrangian multipliers $\omega_i$ satisfying the saddle point equations
\be
k_i=\sum_{j\neq i} \frac{e^{\omega_i^{\star}+\omega_j^{\star}}}{1+e^{\omega_i^{\star}+\omega_j^{\star}}},
\label{sp1}
\ee
{
and the coefficients $\alpha_i $ defined as
\be
\alpha_i\simeq\sum_j\frac{e^{\omega_i^{\star}+\omega_j^{\star}}}{\left(1+e^{\omega_i^{\star}+\omega_j^{\star}}\right)^2},
\ee
}
The probability of a link $i,j$ in this ensemble is given by
\be
p_{ij}^{(1)}=
\frac{e^{\omega_i^{\star}+\omega_j^{\star}}}{1+e^{\omega_i^{\star}+\omega_j^{\star}}}.
\label{hv.eq}
\ee
In particular in this ensemble $p_{ij}\neq f(\omega_i)f(\omega_j)$,
consequently {\it the model retains some ``natural'' correlations} \cite{Garlaschelli} given by the
degree sequence and the constraint that we consider only simple
networks. In fact these are nothing else than the correlations of the
configuration model \cite{Molloy_Reed}.

The "canonical model" corresponding to the configuration model is then
a "hidden variable" models  where each node $i$ is assigned a "hidden
variable" $\omega_i$ and the probability for each link follow
$(\ref{hv.eq})$.
Similar expressions where already derived in different papers \cite{hv1,Garlaschelli,hv4} but with a
different interpretation. Here the "hidden variables'' $\omega_i$ are
simply fixing the average degrees  of each node.
We note here that  the derivation of \cite{Boguna_hv} guarantees that
in the "canonical" model the connectivity of each node is distributed 
according to a Poisson distribution with average $\sum_j p_{ij}$.

The form of the probability $p_{ij}$ is such that when inferring the
values of the "hidden variables" $\omega_i$  for a "canonical" network
in this ensemble by maximum likelihood methods, we obtain the
$\omega^{\prime}_i=\omega_i$ in the large network limit \cite{Garlaschelli}.

\subsubsection{ Uncorrelated networks}
The case in which  
there is a structural cutoff in the network $k_i<\sqrt{\avg{k}N}$ is of particular interest.
In this case we can approximate Eq. $(\ref{sp1})$ by
$e^{\omega_i^{\star}}\simeq k_i/
\sqrt{\avg{k}N},\alpha_i\simeq k_i$.
In this limit the network {\it is uncorrelated}
the probabilities of a link are given by $p_{ij}^{(1),uncorr}=k_ik_j/(\avg{k}N)$,
since the $\omega^*_i<0$. We call the entropy of these uncorrelated
ensembles  the {\it structural entropy} $\Sigma_S$  and we can evaluate it
providing the explicit expression 
\bea 
N\Sigma_S&\simeq&-\sum_i \ln[k_i/\sqrt{\avg{k}N}]k_i
-\frac{1}{2}\sum_i\ln(2\pi k_i)\nonumber \\
&&+\frac{1}{2}\sum_{ij} \frac{k_i
 k_j}{\avg{k}N}-\sum_{ij}\frac{1}{4}\frac{k_i^2 k_j^2}{(\avg{k}N)^2}+\dots\nonumber \\
&= &-\sum_i (\ln k_i-1) k_i-\frac{1}{2}\sum_i \ln(2\pi k_i)+
\nonumber \\
&& \frac{1}{2}\avg{k}N
[\ln(\avg{k}N)-1]
-\frac{1}{4}\left(\frac{\avg{k^2}}{\avg{k}}\right)^2+\dots.
\label{Sunco}
\eea

Expression $(\ref{Sunco})$ gives for the number of networks in the ensemble
\be
{\cal N}_S^{uncorr}\simeq\frac{(\avg{k}N)!!}{\prod_i
 k_i!}\exp\left[-\frac{1}{4}\left(\frac{\avg{k^2}}{\avg{k}}\right)^2+{\cal O}(\ln N)\right].
\label{N_1_unc}
\ee 

From combinatorial arguments we can derive an expression ${\cal N}_C^{uncorr}$ for the
number of uncorrelated networks with a given degree sequence which agrees with the
above estimate $(\ref{N_1_unc})$ in the limit $N\gg 1$, i.e. 
\be
\ln{\cal N}_c^{uncorr}=\ln {\cal N}_S^{uncorr}+{\cal O}(\ln N).
\ee 
In fact by combinatorial arguments we can show that the number of networks with
given degree sequence is given by the following expression in the
large $N$ limit, i.e.
\be
{\cal
N}_c^{uncorr}\propto\frac{(2L-1)!!e^{-\frac{1}{4}\left(\frac{\avg{k^2}}{\avg{k}}\right)^2}}{\prod_ik_i!}
\label{N_SG}
\ee
The factor $(2L-1)!!$ accounts for the total number of wiring's of the
links.
 In fact if we want to construct a network, given a certain
distribution of half-edges through the $N$ nodes of the network, as a
first step we take a half-edge and we match it with one of the $2L-1$
other half-edge of the network. Secondly we match a
new half-edge with one of the $2L-3$ remaining half-edges.
Repeating this procedure we get one out of $(2L-1)!!$ possible
wiring of the links. This number includes also the wiring of the
links which gives rise to networks with double links.
To estimate the number of such undesired wiring we assume that the network is random, i.e. that the probability that a node with $k_i$ half-edges connects to a node with
 $k_j$ half-edges is a Poisson variable with average $k_i k_j/(\avg{k}N)$. In
 this hypothesis the probability $\Pi$ that the network does not
 contain double links is equal to \cite{Chaos}
\be
\Pi=\prod_{i<j} \left(1+\frac{k_i k_j}{\avg{k}N} \right)e^{-\frac{k_i
k_j}{<k>N}}\sim e^{-\frac{1}{4} \left(\frac{\avg{k^2}}{\avg{k}}\right)^2}.
\ee
Finally in the expression $(\ref{N_SG})$ for ${\cal N}_{c}$ there is
an additional term which takes into account the number of wiring of
the links giving rise to equivalent networks without double
links. This term is given by the number of possible permutation of the
half-edges at each node, i.e. $\prod_i k_i!$.
We note here that a similar result was derived by mathematicians for
the case in which the maximal connectivity $K<N^{1/3}$ \cite{Enumeration1} and an
inequality was proved for the case $K>N^{1/3}$
 \cite{Enumeration2}.
Now we extend these results by statistical mechanics methods to
uncorrelated networks with maximal connectivity $K<\sqrt{\avg{k}N}$.

\subsection{ The entropy of a network ensemble with fixed degree
  correlations}

We consider now network ensembles with given degree correlations and
given average degree of neighboring nodes, satisfy the constraints
defined in Eqs. $(\ref{dc1.eq})$ and $(\ref{dc2.eq})$. 
We can proceed to the evaluation of the probability of a link
$p_{ij}^{(2)}$ and the calculation of the entropy of the ensemble as in
the configuration model.
In this case we have to introduce the Lagrangian multipliers
$\omega_i$ fixing the degree of node $i$ and the Lagrangian
multipliers $A_{k}$ fixing the average degree of nodes of degree $k$.

The partition function of this ensemble can be evaluated at the saddle
point  giving for the entropy of the ensemble, in the thermodynamic limit
value
\bea
N\Sigma_2^{und}&\simeq&-\sum_i\omega_i^*k_i-\sum_k A_k^* k_{nn}(k) k N_k\nonumber \\
& &+\sum_{i<j}\ln(1+e^{\omega_i^* +\omega_j^*+k_i A_{k_{j}}^*+k_j
 A_{k_{i}}^*})\nonumber \\
&&-\frac{1}{2}\sum_i \ln(2\pi \alpha_i)-\frac{1}{2}\sum_k \ln(2\pi \alpha_k)
\eea
where $\omega_i^{\star}$ and $A_k^{star}$ satisfy the saddle point equations
\bea
k_i&=&\sum_{j\neq i}
\frac{e^{\omega_i^*+\omega_j^*+k_j A_{k_i}^*+k_i
 A_{k_j}}}{1+e^{\omega_i^*+\omega_j^*+k_j A_{k_i}^*+k_i A_{k_j}^*}},
 \\
k_{nn}(k)&=&\frac{1} {k N_k}\sum_i\delta(k_i-k)\sum_{j\neq i} k_j \frac{e^{\omega_i^*+\omega_j^*+k_j A_{k_i}^*+k_i A_{k_j}^*}}{1+e^{\omega_i^*+\omega_j^*+k_j A_{k_i}^*+k_i A_{k_j}^*}}. \nonumber
\label{sp}
\eea
and where
with $\alpha_i,\alpha_k$ are approximately equal to the following expressions
\bea
\alpha_i&\simeq&\sum_j\frac{e^{\omega_i^*+\omega_j^*+k_j A_{k_i}^*+k_i
 A_{k_j}^*}}{\left(1+e^{\omega_i^*+\omega_j^*+k_j A_{k_i}^*+k_i
 A_{k_j}^*}\right)^2} \\
\alpha_k&\simeq&\sum_i\delta(k_i-k)\sum_{j\neq i} k_j^2 \frac{e^{\omega_i^*+\omega_j^*+k_j A_{k_i}^*+k_i A_{k_j}^*}}{\left(1+e^{\omega_i^*+\omega_j^*+k_j A_{k_i}^*+k_i A_{k_j}^*}\right)^2}.\nonumber
\eea
The probability $p_{ij}^{(2)}$ of the link $(i,j)$ in this ensemble is
given by
\be
p_{ij}^{(2)}=\frac{e^{\omega_i^*+\omega_j^*+k_j A_{k_i}^*+k_i
 A_{k_j}}}{1+e^{\omega_i^*+\omega_j^*+k_j A_{k_i}^*+k_i A_{k_j}^*}}.
\ee
This formula generalize the "hidden variable" formula of the
configuration model to networks with strong degree-degree correlations.
In particular in order to build a "canonical" network with strong
degree degree correlation
we can consider nodes with "hidden variables" $\theta_i$ and
$G_{\theta}$ and a probability $p_{ij}$ to have a link between a node $i$ and
a node $j$ given by
\be
p_{ij}=\frac{\theta_i\theta_j (G_{\theta_i})^{\theta_j}(G_{\theta_j})^{\theta_i}}{1+\theta_i\theta_j (G_{\theta_i})^{\theta_j}(G_{\theta_j})^{\theta_i}}.
\ee

\subsection{The entropy of network ensemble with given degree sequence
  and given community structure}

The partition function of network ensembles with given degree sequence $(\ref{c1.eq})$
and given community structure $(\ref{c2.eq})$ can be evaluated by saddle point
approximation in the large network limit as long as $Q={\cal O}(N^{1/2})$. 

Following the same steps as in the previous case we find that the
entropy for such an ensemble is given by
\bea
N\Sigma_c&\simeq&-\sum_ik_i\omega_i^{\star}-\sum_{q\leq q'}A(q,q')w(q,q')^{\star}\nonumber
\\
&&+\sum_{i<j}\ln\left(1+e^{\omega_i^{\star}+\omega_j^{\star}+w^{\star}({\underline{q}_{ij},\overline{q}_{ij}})}\right)\nonumber\\
&& -\frac{1}{2}\sum_i \ln(2\pi \alpha_i)-\frac{1}{2}\sum_{q<q'}\ln(2\pi\alpha_{q,q'})
\eea
with the Lagrangian multipliers $\{\omega_i^{star}\},\{w_{q,q'}^{star}\}$ satisfying the saddle point equations
\bea
k_i&=&\sum_{j\neq i}\frac{e^{\omega_i^{\star}+\omega_j^{\star}+w^{\star}(\underline{q}_{ij},\overline{q}_{ij})}}{1+e^{\omega_i^{\star}+\omega_j^{\star}+w^{\star}(\underline{q}_{ij},\overline{q}_{ij})}} \\
A(q,q')&=&\sum_{i<j}\delta({\underline{q}_{ij}-q})\delta(\overline{q}_{ij}-q')\times\nonumber
\\
&& \times\frac{e^{\omega_i^{\star}+\omega_j+w^{\star}(q,q')}}{1+e^{\omega_i^{\star}+\omega_j^{\star}+w^{\star}(q,q')}},\nonumber
\eea 
and  with $\alpha_i,\alpha_{q,q'}$ that can be approximated by 
\bea
\alpha_i&\simeq&\sum_j\frac{e^{\omega_i^{\star}+\omega_j^{\star}+w^{\star}(\underline{q}_{ij},\overline{q}_{ij})}}{\left(1+e^{\omega_i^{\star}+\omega_j^{\star}+w^{\star}(\underline{q}_{ij},\overline{q}_{ij})}\right)^2}\\
\alpha_{q,q'}&\simeq&\sum_{i<j}\delta({\underline{q}_{ij}-q})\delta(\overline{q}_{ij}-q')\times\nonumber
\\
&&\times\frac{e^{\omega_i^{\star}+\omega_j^{\star}+w^{\star}(q,q')}}{\left(1+e^{\omega_i^{\star}+\omega_j^{\star}+w^{\star}(q,q')}\right)^2}\nonumber .
\eea
In this ensemble the probability for a link $p_{ij}^{(c)}$ between a
node $i$ and a node $j$ is equal to
\be
p_{ij}^{(c)}=\frac{e^{\omega_i^{\star}+\omega_j^{\star}+w^{\star}(\underline{q}_{ij},\overline{q}_{ij})}}{1+e^{\omega_i^{\star}+\omega_j^{\star}+w^{\star}({\underline{q}_{ij},\overline{q}_{ij}})}}.
\ee
Assigning each node a "hidden variable" $\theta_i$ and to each pair of
communities the symmetric matrix $V(q,q')$ we can construct the
``hidden variable" or "canonical'' ensemble by extracting each link with
probability 
\be
p_{ij}=\frac{\theta_i \theta_j V(q_i,q_j)}{1+\theta_i \theta_j V(q_i,q_j)}
\ee.

\subsection{The entropy of a network ensemble with given distance between
 the nodes}
Finally we consider the ensemble of undirected networks 
 living in a generic embedding space and with structural constraints
 described by $(\ref{d1.eq})$ and $(\ref{d2.eq})$.
Following the same steps as in the previous cases we find that the
entropy for such an ensemble in the large network limit is given by 
\bea
N\Sigma_d&\simeq&-\sum_ik_i\omega_i^{\star}-\sum_{\ell=1}^{\Lambda}B(d_{\ell})g({d_{\ell}})^{\star}\nonumber
\\
&&+\sum_{i<j}\ln\left(1+e^{\omega_i^{\star}+\omega_j^{\star}+\sum_{\ell}\chi_{\ell}(d_{ij})g^{\star}(d_{\ell})}\right)\nonumber\\
&& -\frac{1}{2}\sum_i \ln(2\pi \alpha_i)-\frac{1}{2}\sum_{\ell=1}^{\Lambda}\ln(2\pi\alpha_{\ell})
\eea
with the Lagrangian multipliers $\{\omega_i\},\{g_{d}\}$ satisfying the saddle point equations
\bea
k_i&=&\sum_{j\neq i}\frac{e^{\omega_i^{\star}+\omega_j^{\star}+\sum_{\ell}\chi_{\ell}(d_{ij})g^{\star}(d_{\ell})}}{1+e^{\omega_i^{\star}+\omega_j^{\star}+\sum_{\ell}\chi_{\ell}(d_{ij})g^{\star}(d_{\ell})}} \\
B(d_{\ell})&=&\sum_{i<j}\chi_{\ell}(d_{ij})\frac{e^{\omega_i^{\star}+\omega_j+g^{\star}(d_{\ell})}}{1+e^{\omega_i^{\star}+\omega_j^{\star}+g^{\star}(d_{\ell}))}},\nonumber\eea 
and the variables $\alpha_i,\alpha_{q,q'}$ approximated by the expressions
\bea
\alpha_i&\simeq&\sum_j\frac{e^{\omega_i^{\star}+\omega_j^{\star}+\sum_{\ell}\chi_{\ell}(d_{ij})g^{\star}(d_{\ell})}}{\left(1+e^{\omega_i^{\star}+\omega_j^{\star}+\sum_{\ell}\chi_{\ell}(d_{ij})g^{\star}(d_{\ell})}\right)^2}\nonumber\\
\alpha_{\ell}&\simeq &\sum_{i,j}\chi_d(d_{ij})\frac{e^{\omega_i^{\star}+\omega_j^{\star}+g^{\star}(d_{\ell})}}{\left(1+e^{\omega_i^{\star}+\omega_j^{\star}+g^{\star}(d_{\ell})}\right)^2}
\eea

The probability for a link between node $i$ and $j$ is equal to
\be
p_{ij}^{(d)}=\sum_{\ell}\chi_{\ell}(d_{ij})\frac{e^{\omega_i^{\star}+\omega_j^{\star}+g^{\star}(d_{\ell})}}{1+e^{\omega_i^{\star}+\omega_j^{\star}+g^{\star}{d_{\ell}}}}.
\ee
Therefore the "hidden variable" model associated to this ensemble
correspond to a model where we fix the "hidden variables'' $\theta_i$
and $W(d_{\ell})$ and we draw a link between node $i$ and node $j$ 
according to
\be
p_{ij}=\sum_{\ell}\chi_{\ell}(d_{ij})\frac{\theta_i\theta_j W(d_{\ell})}{1+\theta_i \theta_j W(d_{\ell})}.
\ee

\section{Weighted networks}

Many networks not only have a non trivial topological structure but
are also characterized by weighted links.
We will assume in this paper that the weight of a link 
 can assume only integer values and consequently a link between a node $i$ and 
node $j$ is characterized by an integer  number $a_{ij}\geq 1$, this is not a
very stringent constraints since we can assume to have always finite
networks (studied in the thermodynamic limit).
In a weighted network the degree and the  strength $s_i$ of the node $i$  are defined as 
\bea
k_i&=\sum_{j\neq i}\Theta(a_{ij})\nonumber \\
s_i&=\sum_{j\neq i} a_{ij}
\eea
where $\Theta(x)=0 $ if $x=0$ and $\Theta(x)=1$ is $x>0$.
It is possible to define series of weighted networks by
considering networks with fixed total strength, with given strength
sequence, with given strength and degree sequence and proceeding by
adding additional features as in the unweighted case.
Here an in the following we study the most relevant cases:
\begin{itemize}
\item
{\it }
We first consider the network ensemble with given total strength $S$.
The structural constraint in this case is equal to 
\be
F({\bf a})-C=\sum_{i<j}a_{ij}-S=0.
\ee 
\item 
{\it ii)}
We consider the network with given strength sequence ${s_1,\ldots,
  s_N}$.
The structural constraints are for this ensemble given by
\be
F({\bf a})_{\alpha}-C_{\alpha}=\sum_{j}a_{\alpha j}-s_{\alpha}=0.
\ee 
for $\alpha=1,\ldots, N$.
\item
{iii)}
Finally we consider the network ensemble with given degree sequence $\{k_1,\ldots,k_n\}$
and strength sequence $\{s_1,\ldots,s_N\}$.
For this ensemble the structural constraints are given by
\be
F({\bf a})_{\alpha}-C_{\alpha}=\sum_{j}\Theta(a_{\alpha j})-k_{\alpha}=0.
\ee 
for $\alpha=1,\ldots, N$ and 
\be
F({\bf a})_{\alpha}-C_{\alpha}=\sum_{j}a_{\alpha j}-s_{\alpha}=0.
\ee 
for $\alpha=1,\ldots, 2N$.
\end{itemize}
\subsection{The entropy of weighted network ensembles with given
 total strength $S$}
The entropy of this ensemble is given by
\be
N\Sigma_1^{W}= \ln\left[\left(\begin{array}{c} \frac{N(N-1)}{2}+S\nonumber
 \\ \frac{N(N-1)}{2} \end{array}\right)\right]
\ee
The average value of a the weight of the link from $i$ to $j$ is given by
\be
w_{ij}=\avg{a_{ij}} _{1}^W=\frac{S}{\frac{N(N-1)}{2}}
\ee
and the probability of a link between node $i$ and $j$ is equal to 
\be
p_{ij}^{W,1}=\frac{S}{S+\frac{N(N-1)}{2}}.
\ee
Therefore the simple networks with adjacency matrix $((A_{ij}))$ that can be constructed from the weighed
networks with adjacency matrix $((a_{ij}))$ by putting
$A_{ij}=\Theta(a_{ij})$ $ \forall i,j$ is uncorrelated.
The canonical ensemble is given by Eq. $(\ref{CW})$
with 
\be
\pi_{ij}(a_{ij})=\frac{e^{\omega a_{ij}}}{1-e^{\omega}}
\ee
and $\omega=-\ln[1+N(N-1)/(2S)]$.

\subsection{The entropy of weighted network ensembles with given strength sequence}
To calculate the entropy of undirected networks with a given strength
sequence of degrees $\{s_i\}$ we proceed by the saddle point
approximation as in previous cases
We find that the entropy of this ensemble of networks is given by 
\be
N\Sigma_1^{W}\simeq-\sum_i\omega_i^{\star} s_i-\sum_{i<
 j}\ln(1-e^{\omega_i^{\star}+\omega_j^{\star}})-\frac{1}{2}\sum_i \ln(2\pi \lambda_i)
\ee
with the Lagrangian multipliers $\omega_i^{\star}$ satisfying the saddle point equations
\bea
s_i&=&\sum_{j\neq i}
\frac{e^{\omega_i^{\star}+{\omega}_j^{\star}}}{1-e^{\omega_i^{\star}+{\omega}_j^{\star}}}.
\label{spS}
\eea
and with $\lambda_{i}$ being the eigenvectors of the Jacobian of the
function
\be
{\cal F}=\sum_{i<
 j}\ln\left[1-e^{-\omega_i-{\omega}_j}\right].
\ee
The average value of a the weight of the link from $i$ to $j$ is given by
\be
\avg{a_{ij}}_{1}^W=\frac{e^{\omega_i^{\star}+{\omega}_j^{\star}}}{1-e^{\omega_i^{\star}+{\omega}_j^{\star}}}.
\ee
and the probability of a link between node $i$ and $j$ is equal to 
\be
p_{ij}^{W,1}=e^{\omega_i^{\star}+\omega^{\star}_j}.
\ee
Therefore as it has been observed in \cite{rc} only by rewiring the
links of a network allowing for multilinks we get a network structure
which is uncorrelated.

The canonical ensemble $(\ref{CW})$ in this case can be constructed by assigning to
every possible link  $(i,j)$ the weight $a_{ij}$ with the probability
\be
\pi_{ij}(a_{ij})=\frac{e^{(\omega_i^{\star}+\omega_j^{\star})a_{ij}}}{1-e^{\omega_i^{\star}+{\omega}_j^{\star}}}.
\ee

\subsection{The entropy of weighted network ensembles with given strength /degree sequence}
The entropy of weighted  networks with a given strength
and degree sequence $\{s_i,k_i\}$ in the large size network limit is
given by  
\bea
N\Sigma_2^{W}&=&-\sum_i\omega_i^{\star} s_i-\sum_i \psi_i^{\star}k_i-\sum_i\nonumber\\
&& +\sum_{i<
 j}\ln\left[1+e^{\psi_i^{\star}+{\psi}_j^{\star}}\frac{1}{e^{-\omega_i^{\star}-
 {\omega}_j^{\star}}-1}\right]\nonumber \\
&& +\frac{1}{2}\sum_{\ell=1}{2N}\sum_i\ln(2\pi \lambda_{\ell})
\eea
with the Lagrangian multipliers satisfying the saddle point equations
\bea
k_i&=&\sum_{j\neq i}
\frac{e^{\psi_i^{\star}+{\psi}_j^{\star}}}{e^{\psi_i^{\star}+{\psi}_j^{\star}}+e^{-(\omega_i^{\star}+ {\omega}_j^{\star})}-1}.\nonumber
\\
s_i&=&\sum_{j\neq i}
\frac{e^{-(\omega_i^{\star}+{\omega}_j^{\star})+(\psi_i^{\star}+{\psi}_j^{\star})}}{(e^{\psi_i^{\star}+{\psi}_j^{\star}}+e^{-(\omega_i^{\star}+ {\omega}_j^{\star})}-1)(e^{-\omega_i^{\star}- {\omega}_j^{\star}}-1)}\\
\label{sp2}
\eea
and with $\lambda_{\ell}$ being the eigenvectors of the Jacobian of the
function
\be
{\cal F}=\sum_{i<
 j}\ln\left[1+e^{\psi_i+{\psi}_j}\frac{1}{e^{-\omega_i-
 {\omega}_j}-1}\right]
\ee
calculated at the values ${\{\omega_i^{\star},\psi^{\star}_i\}}$.
The average weight of the link $(ij)$ is given by 
\be
\avg{ a_{ij}}_{2}^W=
\frac{e^{-(\omega_i^{\star}+{\omega}_j^{\star})+(\psi_i^{\star}+{\psi}_j^{\star})}}{(e^{\psi_i^{\star}+{\psi}_j^{\star}}+e^{-(\omega_i^{\star}+ {\omega}_j^{\star})}-1)(e^{-\omega_i^{\star}- {\omega}_j^{\star}}-1)}
\ee
and the probability of a link between node $i$ and $j$ is equal to 
\be
p_{ij}^{W,2}=
\frac{e^{\psi_i^{\star}+{\psi}_j^{\star}}}{e^{\psi_i^{\star}+{\psi}_j^{\star}}+e^{-(\omega_i^{\star}+ {\omega}_j^{\star})}-1}
\ee

The canonical ensemble $(\ref{CW})$ in this case con be constructed by assigning to
every possible link  $(i,j)$ the weight $a_{ij}$ with the probability
\be
\pi_{ij}(a_{ij})=\frac{e^{(\psi_i^{\star}+\psi_j^{\star})\Theta(a_{ij})}e^{(\omega_i^{\star}+\omega_j^{\star})a_{ij}}}{1+e^{\psi_i^{\star}+{\psi}_j^{\star}}\frac{1}{e^{-\omega_i^{\star}-
 {\omega}_j^{\star}}-1}}.
\ee

\section{ Directed networks}
An undirected network is determined by a symmetric adjacency
matrix, while the matrix of a directed network is in general
non-symmetric.
Consequently the degrees of freedom of a directed network are more
 than the degrees of freedom of an undirected network.
In the following we only consider the network ensemble with 
\begin{itemize}
\item
{\it i)} Total number of directed links
The structural constraint in this case is equal to
\be
F({\bf a})-C=\sum_{ij}a_{ij}-S=0.
\ee
\item
{\it ii)} Given directed degree sequence $\{k_1^{in},
k_1^{out},\ldots, k_N^{(in)},k_N^{(out)}\}$.
The structural constraints in this case are 
\be
F({\bf a})_{\alpha}-C_{\alpha}=\sum_{j}a_{\alpha j}-k_{\alpha}^{out}=0.
\ee 
for $\alpha=1,\ldots, N$ and 
\be
F({\bf a})_{\alpha}-C_{\alpha}=\sum_{j}a_{j \alpha }-k_{\alpha}^{in}=0.
\ee 
for $\alpha=N+1,\ldots, 2N$.
\end{itemize}

\subsection{The entropy of directed network ensembles with fixed number of directed links }
If we consider the number of directed networks ${\cal N}_0^{dir}$ with
given number of nodes and of directed links we find
\bea
{\cal N}_0^{dir}=\left(\begin{array}{c} N(N-1) \\ L^{dir}\end{array}\right).
\eea
In this case the probability of a directed link is given  by
\be
p_{ij}=\frac{L}{N(N-1)}.
\ee
\subsection{The entropy of directed network ensembles with given degree sequence}
To calculate the entropy of directed networks with a given degree sequence of
in/out degrees $\{k_i^{out},k_i^{in}\}$ we just have to impose 
the constraints on the incoming and outgoing connectivity,
\bea
Z_1^{dir}&=&\sum_{\{a_{ij}\}}\prod_i\delta(k_i^{(out)}-\sum_j a_{ij})
\prod_i \delta(k_i^{(in)}-\sum_j a_{ji})\nonumber \\
& &\exp[\sum_{ij}h_{i,j}a_{ij}]
\eea 
Following the same approach as for the undirected case, { we
 find that the entropy of this ensemble of networks is given by 
\bea
N\Sigma_1^{dir}&\simeq&-\sum_i\omega_i^{\star} k_i^{(out)}-\sum_i k_i^{(in)}
\hat{\omega}_i^{\star}\nonumber \\
& & +\sum_{i\neq j}\ln(1+e^{\omega_i^{\star}+ \hat{\omega}_j^{\star}})\nonumber \\
&&-\frac{1}{2}\sum_i \ln((2\pi)^2 \alpha^{(in)}_i \alpha^{(out)}_i)
\eea
}
with the Lagrangian multipliers satisfying the saddle point equations
\bea
k_i^{(out)}&=&\sum_{j\neq i}
\frac{e^{\omega_i^{\star}+\hat{\omega}_j^{\star}}}{1+e^{\omega_i^{\star}+\hat{\omega}_j^{\star}}}.\nonumber
\\
k_i^{(in)}&=&\sum_{j\neq i}
\frac{e^{\omega_j^{\star}+\hat{\omega}_i^{\star}}}{1+e^{\omega_j^{\star}+\hat{\omega}_i^{\star}}}.
\label{sp3}
\eea

with 
\bea\alpha^{(out)}_i&\simeq&\sum_{j\neq i}
\frac{e^{\omega_i^{\star}+\hat{\omega}_j^{\star}}}{(1+e^{\omega_i^{\star}+\hat{\omega}_j^{\star}})^2}\nonumber \\
\alpha_i^{(in)}&\simeq&\sum_{j\neq i}
\frac{e^{\omega_j^{\star}+\hat{\omega}_i^{\star}}}{(1+e^{\omega_j^{\star}+\hat{\omega}_i^{\star}})^2}
\eea

The probability for a directed link from $i$ to $j$ is given by
\be
p_{ij}^{(1,dir)}=\frac{e^{\omega_i^{\star}+\hat{\omega}_j^{\star}}}{1+e^{\omega_i^{\star}+\hat{\omega}_j^{\star}}}.
\ee

If the $\omega_i+\hat{\omega}_j<0 \forall i,j=1,\dots N$ the directed
network becomes uncorrelated and we have
$p_{ij}^{1,(dir)}=k_i^{(out)}k_j^{(in)}/\sqrt{\avg{k_{in}}N}$.
Given this solution the condition for having uncorrelated directed
networks is that the maximal in-degree $K^{(in)}$ and the maximal
out-degree $K^{(out)}$ should satisfy,
$K^{(in)}K^{(out)}/\sqrt{\avg{k_{in}}N}<1$.
The entropy of the directed uncorrelated network is then given by 
{
\bea
N\Sigma_{1,dir}^{uncorr}&\simeq& \ln(\avg{k_{in}}N)!-\sum_i \ln(k_i^{(in)}!k_i^{(out)}!)\nonumber \\
&&-\frac{1}{2}\frac{\avg{k_{in}^2}}{\avg{k_{in}}}\frac{\avg{k_{out}^2}}{\avg{k_{out}}}
\eea 
}
which has a clear combinatorial
interpretation as it happens also for the undirected case.

\section{ Natural degree distribution corresponding to a given  structural entropy}

 For power-law networks with power-law exponent $\gamma\in(2,3)$ the
 entropy of the networks with fixed degree sequence  $\Sigma_1$ given by Eq. $(\ref{Sigma_1})$ decreases with the value of the
power-law exponent $\gamma$ when we compare network ensemble  with the
same average degree \cite{entropy}.
Therefore scale-free networks have much smaller entropy than
homogeneous networks. This fact seems to be in contrast with the fact
that scale-free networks are the underlying  structure of a large
class of complex systems.
The apparent paradox can be easily be resolved if we consider that
many networks are the result of a non-equilibrium dynamics. Therefore
they do not have to satisfy the maximum entropy principle.
Nevertheless, in  order to give more insight  and comment on  the universal occurrence of power-law networks 
in this section we derive the most likely degree distribution of
given {\it structural entropy} when the total number of nodes and links are
kept fixed. 
By structural entropy we define the entropy $\Sigma_S$ $(\ref{Sunco})$ of
uncorrelated networks with fixed degree distribution. 
In order to do that we construct a statistical model very closely
related to the urn or ``ball in the box'' models \cite{Ritort,Doro_stat}

We consider degree distributions $\{N_k\}=\sum_i \delta(k-k_i)$ which arise from the random
distribution of the $2L$ half-edges through the $N$ nodes of the network.
The number of ways ${\cal N}_{\{N_k\}}$ in which we can distribute
the $(2L)$ half-edges in order to have a $\{N_k\}$ degree
distribution are 
\be
{\cal N}_{\{N_k\}}=\frac{(2L)!}{\prod_k (kN_k)!}.
\ee
We want to find the most likely degree distribution that corresponds to a given value of the structural entropy.

Proceeding as in standard statistical mechanics, we define a
normalized partition function ${\cal Z}$ as
 \be {\cal Z}=\frac{1}{C}\sum_{ \{N_k \} }' {\cal{N}}_{ \{N_k \}} e^{\beta
N\Sigma_S(\{N_k\})}.
\label{Z.eq} \ee with $C=(2L)!\exp[\beta (2L)!!]$.
The role of the parameter $\beta$ in Eq. $(\ref{Z.eq})$ is to fix the average value of the
structural entropy $\Sigma_S$. When $\beta\rightarrow \infty$ the
structural entropy $\Sigma_S$
is maximized when $\beta \rightarrow \beta_{min}$ the structural entropy $\Sigma_S$
is minimized.

In equation $(\ref{Z.eq})$ the sum $\sum'$ over the $\{N_k\}$
distributions is extended only to $\{N_k\}$ for which the total number
of nodes $N$ and the total number of links $L$ in the network is
fixed, i.e. \bea \sum_k N_k=N \nonumber \\ \sum_k k N_k=2 L.
\label{conditions} \eea To enforce these conditions we introduce in
$(\ref{Z.eq})$ the delta functions in the integral form providing the
expression 
\begin{widetext}
\bea {\cal Z}&=&\frac{1}{(2L)!}\int \frac{d\lambda}{2\pi} \int dS
\int \frac{d\mu}{2\pi}\int \frac{d\nu}{2\pi} \sum_{ \{N_k \} }
\exp\left[ -\beta \sum_k N_k \ln k! -\frac{\beta}{4}
\left(\frac{S}{\avg{k}}\right)^2-\sum_k \ln[(kN_k)!]\right.\nonumber
\\ & &\left.-i\lambda (2L-\sum_k N_k k)-i\mu (N-\sum_k N_k)-i\nu(N
S-\sum_k k^2 N_k)\right].\eea
\bea {\cal Z}&=&\int d S\int
\frac{d\lambda}{2\pi} \int \frac{d\mu}{2\pi} \int \frac{d\nu}{2\pi}
\exp\left[-i\lambda 2L -i\mu N-i\nu N
S-\frac{\beta}{4}\left(\frac{S}{\avg{k}}\right)^2+ \sum_k \ln
G_k(\lambda,\mu,\nu)\right]=\nonumber \\ &=&\int dS \int
\frac{d\lambda}{2\pi} \int \frac{d\mu}{2\pi}\int \frac{d\nu}{2\pi}
\exp[Nf(\lambda, \mu,\nu,S)]
\label{Z2} \eea 
where 
\bea
G_k(\lambda, \mu,\nu)&=&\sum_{N_k} \frac{1}{(k
N_k)!} \left\{k N_k\left[i\lambda +i \frac{\mu}{k}+i\nu
k-\frac{\beta}{ k}\ln(k!) \right]\right\}. 
\eea
\end{widetext}
Assuming that the sum
over all $N_k$ can be approximated by the sum over all $L_k=k
N_k=1,2,\dots \infty$ we get $\ln G_k(\lambda,
\mu,\nu)=\exp\left[{i\lambda +i\mu/k -\frac{\beta}{k}\ln(k!)}+i\nu
k\right]$ and 
\bea f(\lambda, \mu,\nu,S)&=&-i\avg{k}\lambda-i\mu-i\nu
S-\frac{\beta}{4}\left(\frac{S}{\avg{k}}\right)^2\nonumber \\
&&+\frac{1}{N}\sum_k
e^{i\lambda+i\mu/k -\frac{\beta}{k}\ln(k!)+i\nu k} \eea where
$<k>=2L/N$ indicates the average degree of the network. By evaluating
$(\ref{Z2})$ at the saddle point, deriving the argument of the
exponential respect to $\lambda$ and $\nu$, we obtain \bea
1=\frac{1}{N}\sum_k \frac{1}{k} e^{i\lambda+i\mu/k
-\frac{\beta}{k}\ln(k!))+i\nu k}. \nonumber \\
\avg{k}=\frac{1}{N}\sum_k e^{i\lambda+i\mu/k
-\frac{\beta}{k}\ln(k!))+i\nu k}\nonumber \\
S=\frac{1}{N}\sum_k
k^2 e^{i\lambda+i\mu/k -\frac{\beta}{k}\ln(k!)+i\nu
  k}\nonumber \\
i\nu N&=&-\beta \frac{S}{2\avg{k}^2}.
\label{dOmega} \eea
 These equations always
have a solution for sparse networks with $L={\cal O}(N)$ provided that
$\beta>1$ and $\avg{k}>1$. The marginal probability that $L_k=k N_k$
is given by \bea P(L_k=kN_k)&=&\frac{1}{(k N_k)!}e^{-\beta
N_k(\ln(k!)+i\lambda k +i\mu +i\nu k^2)}\nonumber \\
&&\times\frac{{\cal Z}_k(L,k N_k,N)}{{\cal Z}(L)},
\label{marginal} \eea with \be {\cal Z}_k(L,\ell,N)=\int dS \int
\frac{d\lambda}{2\pi} \int \frac{d\mu}{2\pi} \int
\frac{d\nu}{2\pi}\exp[Nf_k(\lambda, \mu,\nu,S,\ell)] \ee and
\bea 
f_k(\lambda,\mu,\nu,\ell)&=&-i(\avg{k}-\ell/N)\lambda-i\mu(1-\ell/(kN))+\nonumber
\\ &&\hspace*{-15mm}-i\nu(S-k\ell/N)-\frac{\beta}{2}\left(\frac{S^2}{\avg{k}}+\right)^2\frac{1}{N}\nonumber \\
&& \hspace*{-15mm} + \ln\left[\sum_{s\neq
k} \frac{1}{(sN_s)!}\exp[sN_s[{i\lambda+i\mu/s +i\nu s-\frac{\beta}{s}\ln(s!)}]\right]
\eea
 If we
develop $(\ref{marginal})$ for $\ell \ll L $ and we use the Stirling
approximation for factorials, we get that each variable $L_k$ is a
Poisson variable with mean $\avg{L_k}$ satisfying \be
\frac{\Avg{L_k}}{k}=\Avg{N_k}\simeq
k^{-\beta-1}e^{i\lambda+\beta+i\mu/k+i\nu k}
\label{PL} \ee where we assume that the minimal connectivity of the
network is $k>0$. The average $\Avg{N_k}$ is a power-law distribution
with a lower and upper  effective cutoffs $-i\mu$  and $1/(i\nu)$
fixing the average degree $\avg{k}$, with the   Lagrangian
parameter $\lambda$ fixing the normalization constant and finally  $\beta$
fixing the structural entropy. The distribution of $P(N_k)$ is
finally \be P(N_k)=\frac{k}{(k N_k)!}e^{-\beta N_k\ln(k!)+i\lambda
kN_k+i\mu N_k +i\nu (k)^2} \label{PNKT}\ee

In the limit $\beta \rightarrow \infty$ $(\ref{PNKT})$ is extremely
peaked around  the average degree $k\simeq k^{\star}={\cal O}(\avg{k})$ of
the network and the degree distribution $N_k$ decays at large value of
$N_k$ as a Poisson distribution, i.e. \be P(N_{k})\simeq
\frac{1}{(kN_{k})!}e^{ kN_{k}\left[-\beta\ln(k^{\star}!)/k^{\star}+i\lambda
k^{\star}+ i\mu/k^{\star}+i\nu k^{\star}\right]}.
\label{Pnk} \ee. Therefore for $\beta \rightarrow \infty$ the network
is Poisson like. In the opposite limit of small structural entropy and $\beta$
small  the $P(N_k)$ distribution $(\ref{PNKT})$ develops a fat tail decaying
like a power-law $(\ref{PL})$ with an exponent $\gamma=\beta+1$.
Therefore the natural distribution with a small value of the
structural entropies are decaying as a power law and and smaller
values of the power-law
exponent  correspond to a smaller value of the structural entropy.
When the value of the entropy is minimal,
$\beta\rightarrow 1$ the degree distribution $(\ref{Pnk})$ has a large
tail with an exponent $\gamma\rightarrow 2$.

\section{Conclusions}

In conclusion we have shown that there is a wide set of network
ensembles that can be naturally described by statistical mechanics
methods.
The statistical mechanics method provides the theoretical estimation
of the entropy of these ensembles that  quantify the cardinality of the
network ensembles. We believe that the entropy of randomized ensembles
constructed from a given real networks will be of great applicability
for inference problems defined on technological social and biological
networks.
In this paper we have focused on some theoretical problems that can be
approached with the use of this quantity.
First we have formulated a series of ``canonical'' or ``hidden
variables''  models that can be used for generating networks with
community structure and spatial embedding.
Secondly we have focused on the degree distribution of network. The degree
distributions are not all equivalent. In fact the associated structural
entropy depends strongly on the distribution. In particular the power-law
degree distribution with exponent $\gamma$ and fixed average degree
are associated to a structural entropy that decreases with $\gamma$.
Nevertheless we have shown that power-law degree distributions are the more likely distributions associated to small 
structural entropy. This shed light on the evidence that power-law
networks constitute a large universality class in complex networks
with a non trivial level of organization.

\acknowledgments 
This work was supported by IST STREP GENNETEC
contract No. 034952.

\end{document}

In the study of complex networks \cite{Latora,Dorogovtsev} the extention of the
concept of
entropy \cite{Burda_en,Chaos,entropy,Latora2} has been recently shown
to have many potential applications. 
It is indeed possible to define 
the {\it entropy of an ensemble} 
of networks as the normalized logarithm of the number of
networks in a given ensemble \cite{entropy}.
Given a real network, different types of randomized network ensembles
can be contructed, each one of them retaining structural information at
different levels (degree sequence, degree correlations, community structure,
distance in
embedding space).
Moreover it is possible to define a {\it structural
entropy}, i.e. the entropy of an uncorrelated network ensemble with
given degree sequence. 
This entropy is the entropy of the configuration ensemble
 \cite{Molloy_Reed} and is the normalized logarithm of the number of possible
networks spanned by the algorithm of reshufling links proposed by
Maslov and Sneppen \cite{Sneppen}.
In \cite{entropy} it was shown that the structural entropy of
power-law networks is a incresing function of the exponent
$\gamma\in(2,3)$ of their degree distribution.
Another definition regards the {\it entropy rate} of a diffusion
process in a network, that carries
infomation about the degree distribution and the degree correlations
present in it \cite{Latora2}.

Randomized network ensembles are the null models of real
 networks and are extensively used to compare a real system to a null
 hypothesis. 
In this paper we study network ensembles with the same degree
distribution, the same degree-correlations {and} the same community
structure of any given real network. We characterize these randomized network
ensembles by their 
entropy, i.e. the normalized logarithm of the total number of networks
which are part of these ensembles.
 We estimate the entropy of randomized
ensembles starting from a large set of real directed and undirected
networks. We propose entropy as an indicator to assess
the role of each structural feature in a given real network.We observe that the ensembles with fixed scale-free degree distribution
have smaller entropy than the ensembles with homogeneous degree
distribution indicating a higher level of order in scale-free
networksIn figure \ref{SF.fig} we plot the entropy of a scale-free network
with natural cutoff and fixed average connectivity $\avg{k}=6,8,10$.
The entropy $\Sigma_1$ of the configuration model is decreasing with
decreasing power-law exponent $\gamma$ reaching its minimum at
$\gamma\rightarrow 2$.
This indicates that scale-free networks with low value of $\gamma$
presents higher level of ordering with respect to random homogeneous
networks.

The complex structural organization \cite{BarabasiNewman} of a real
network is intimately related with the
dynamical process that takes place on it.
Real networks show different levels of 
organization. 
To characterize the structure of undirected networks few different quantities have been
proposed: (i)
the density of the links, (ii) the degree sequence \cite{BA}, (iii)
the degree-degree correlations \cite{Vespignani_corr,Sneppen,Berg}, (iv) the
clustering coefficient \cite{SW,Modular}, (v) the k-core structure \cite{k-core_kirk,k-core_doro,k-core_vesp}
and (vi) the community structure \cite{Newman1,Danon,Newman2,review}
and finally the nature of the embedding space
 \cite{Kleinberg,Dorogovtsev_new, Boguna2}.
For weighted networks very informative are strength-degree
correlations \cite{Vepsignani_weigh}
and for directed networks in-degree, out-degree correlations \cite{lood_adilson}. 
To characterize the complexity of a real network we consider a series of
randomized network models which retain an increasing number of
structural features of the
real network. As the number of features shared by the random ensemble with the real
network increase, the number of networks in the random ensemble
will decrease.
In \cite{entropy} we introduced {\it the entropy of an 
 ensemble} as the normalized logarithm of the number of networks in
the ensemble. 
The entropy difference between a random ensemble and a random
ensemble having just one additional structural constraint, can be considered as a indicator to assess
the role of the new structural constraint in a given real network.

 In our approach we will first consider a particular
real network to be part of the 
ensemble of networks with the same number of nodes $N$ and links $L$ the real
network has.
This network ensemble is the $G(N,L)$ studied by the random
graph community.
Subsequently we consider the configuration model of networks with given
degree sequence and we restrict the number of possible networks.
Furthermore we consider the ensemble of networks with a given degree 
sequence and with given degree correlations or with given community structure and we
further restrict the space of possible networks.
We will also consider the ensemble of networks with given community
structure and degree sequence and with given distance between the
nodes living in some embedding space. 
Then we will consider weighted networks with given distribution of the
strength or with given strength-degree distribution.
Finally we will study directed networks with given in-out degree
distribution.

The randomized network ensemble that we introduce, out of different
statistical mechanics network models that have been proposed \cite{Burda_stat,Doro_stat},
is very closely related with the hidden variable model introduced by
 \cite{hv1,hv2,hv3,hv4,Park,Caldarelli,Garlaschelli} 
in which hidden variables $\theta_i$ are associated to each node 
$i$ of the network.

In particular the randomized network ensembles that we construct in
this paper are of two types: the "microcanonical'' 
ensembles composed of networks that have some specific
structural feature, and the "canonical" ensembles (or ``hidden variable''
ensembles) composed of networks
having "typically" (with high probability but not always) the required structural features.

In the second part we will consider the degree distribution arising 
 when we distribute randomly half-edges (or stub) through the nodes of the network
 by keeping fixed the structural entropy. 
In this case the degree distribution that maximizes the
entropy decays as a Poisson distribution.
The degree distribution that minimizes the entropy instead has a tail
which decays as a power-law with an exponent $\gamma\rightarrow 2$.
This result indicates that scale-free degree distributions emerge
naturally when considering networks ensemble with small structural
entropy.
The appearance of the power-law degree distribution reflects the tendency of social, technological and especially biological networks toward ``ordering''.
This tendency is at work regardless of the mechanism which is
driving their evolution that can be either a preferential attachment mechanism \cite{BA}, or a ``hidden variables'' mechanism \cite{hv1,hv2,hv3,hv4,Boguna_hv} or some other statistical mechanics mechanisms \cite{Burda_stat,Doro_stat}.